\documentclass[aps,pre,showpacs,noshowkeys,amsmath,amssymb,amsfonts,superscriptaddress,longbibliography,reprint]{revtex4-1}
\usepackage[english]{babel}

\usepackage{graphicx}
\usepackage{bm}
\usepackage{physics}
\usepackage{mathtools}
\usepackage{gensymb}

\setcitestyle{super}
\usepackage{caption}
\usepackage{subcaption}
\DeclareCaptionLabelSeparator{bar}{~\rule[-0.4ex]{0.2ex}{1em}~}
\DeclareCaptionLabelFormat{subfor}{\textbf{#2}}
\newcommand*\bfcaption[2]{\caption[#1]{\textbf{#1.}#2}}
\usepackage{xcolor}
\definecolor{UBcolor}{HTML}{007CC1}
\usepackage[colorlinks=true,pdfnewwindow=true,linkcolor=UBcolor,citecolor=UBcolor,urlcolor=UBcolor,breaklinks=true,linktocpage]{hyperref}
\usepackage[all]{hypcap}
\usepackage[nameinlink,capitalise]{cleveref}
\crefname{SI section}{SI Section}{SI Sections}
\Crefname{SI section}{SI Section}{SI Sections}
\begin{document}

\title{Fingering Instability of Active Nematic Droplets}

\author{Ricard Alert}
\email{ralert@pks.mpg.de}
\affiliation{Max Planck Institute for the Physics of Complex Systems, N\"{o}thnitzerst. 38, 01187 Dresden, Germany}
\affiliation{Center for Systems Biology Dresden, Pfotenhauerst. 108, 01307 Dresden, Germany}

\date{\today}

\begin{abstract}
From the mitotic spindle up to tissues and biofilms, many biological systems behave as active droplets, which often break symmetry and change shape spontaneously. Here, I show that active nematic droplets can experience a fingering instability. I consider an active fluid that acquires nematic order through anchoring at the droplet interface, and I predict its morphological stability in terms of three dimensionless parameters: the anchoring angle, the penetration length of nematic order compared to droplet size, and an active capillary number. Droplets with extensile (contractile) stresses and planar (homeotropic) anchoring are unstable above a critical activity or droplet size. This instability is interfacial in nature: It arises through the coupling of active flows with interface motion, even when the bulk instability of active nematics cannot take place. In contrast to the dynamic states characteristic of active matter, the instability could produce static fingering patterns. The number of fingers increases with activity but varies non-monotonically with the nematic penetration length. Overall, these results can help to understand the self-organized shapes of biological systems, and to design patterns in active materials.
\end{abstract}

\maketitle

Active matter is driven internally by its own constituents, be they molecular motors, cells, animals, or artificial self-propelled particles. As a result, active fluids exhibit striking phenomena such as spontaneous flows without external driving \cite{Ramaswamy2010,Marchetti2013}, turbulence at low Reynolds numbers \cite{Alert2022}, and phase separation of repulsive particles \cite{Cates2015}. These distinctive phenomena arise from activity-induced bulk instabilities.

\begin{figure}[tb!]
\begin{center}
\includegraphics[width=\columnwidth]{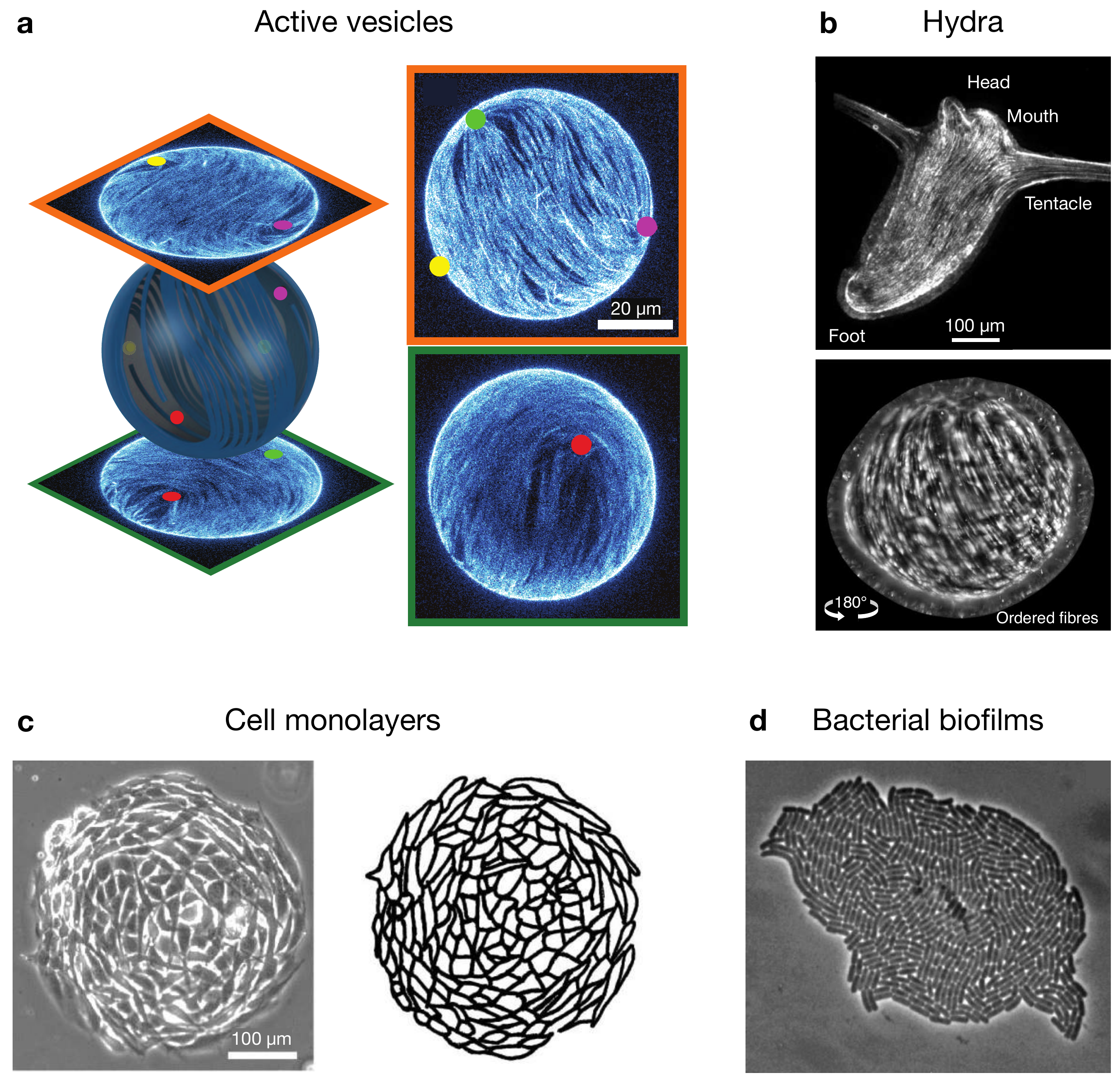}
\end{center}
  {\phantomsubcaption\label{Fig active-vesicles}}
  {\phantomsubcaption\label{Fig hydra}}
  {\phantomsubcaption\label{Fig cell-monolayers}}
  {\phantomsubcaption\label{Fig bacterial-biofilms}}
\bfcaption{Examples of active nematic droplets}{ \subref*{Fig active-vesicles}, An active nematic film made of microtubules and molecular motors is encapsulated within a lipid vesicle. The active film undergoes spontaneous flows, which drive vesicle shape deformations. Dots indicate topological defects of the nematic order. The two images on the right are top and bottom hemisphere projections obtained through confocal microscopy. Adapted with permission from Ref. \cite{Keber2014}. \subref*{Fig hydra}, A mature \textit{Hydra}, a small aquatic animal, has supracellular actin fibers that form an active nematic film in the external cell layer (top). Strikingly, the animal can regenerate from a tissue fragment that initially folds into a closed spheroidal shell (bottom). Adapted with permission from Ref. \cite{Maroudas-Sacks2021}. \subref*{Fig cell-monolayers}, Phase-contrast image (left) and cell boundaries (right) of a circular monolayer of 3T3 fibroblasts. Cells align parallel to the tissue boundary, which induces nematic order that is maximal at the edge and decreases toward the center. Adapted with permission from Ref. \cite{Xie2021}. \subref*{Fig bacterial-biofilms}, A growing biofilm of \textit{E. coli} bacteria. Cells tend to align parallel to the boundary and with one another, and growth-induced active nematic forces drive biofilm shape changes. The image has no scale bar, but the average cell length is $\sim 2-3$ $\mu$m. Adapted with permission from Ref. \cite{Dellarciprete2018}.}
\label{Fig1}
\end{figure}

Very often, however, active fluids form finite droplets. Examples abound in biological systems (\cref{Fig1}), including biomolecular condensates \cite{Weber2019a}, the mitotic spindle \cite{Brugues2014a,Oriola2020,Oriola2018}, cell aggregates and monolayers \cite{Perez-Gonzalez2019,Alert2020,Alert2021d}, and bacterial biofilms \cite{Dellarciprete2018}. Active droplets can also be made artificially \cite{Needleman2017}, for example by preparing vesicles containing either microswimmers \cite{Takatori2020a,Ramos2020,Vutukuri2020,Rajabi2021}, an actomyosin cortex \cite{Carvalho2013,Loiseau2016}, or microtubule-kinesin films \cite{Sanchez2012,Keber2014,Guillamat2018,Chen2021a} (\cref{Fig active-vesicles}). In all these systems, active droplets are commonly observed to spontaneously break symmetry and undergo shape changes. Interestingly, these shape dynamics could provide basic mechanisms for the onset of cell motility \cite{Ziebert2016,Cates2018,Callan-Jones2008,BenAmar2011b,Tjhung2012,Ziebert2012,Blanch-Mercader2013,Whitfield2014,Tjhung2015,Khoromskaia2015,Whitfield2016,Lavi2020,Loisy2020,Stegemerten2021} as well as cell and tissue morphogenesis \cite{Al-Izzi2021,Mietke2019a,Mietke2019b,Maroudas-Sacks2021,Fernandez2021,Khoromskaia2021,Hoffmann2021} (\cref{Fig hydra}).


A key feature of active droplets is that they have an interface. As in passive fluids, interfaces have important consequences such as setting the kinetics of phase separation \cite{Cates2018,Weber2019a,Fausti2021}, driving wetting phenomena \cite{Joanny2012,Perez-Gonzalez2019}, and allowing for interfacial instabilities. Here, I predict a generic interfacial instability of active nematic fluids. Thus, the results provide an active counterpart to paradigmatic interfacial instabilities in passive fluids, such as the Saffmann-Taylor instability underlying viscous fingering \cite{Saffman1958,Casademunt2004}.

In active and living fluids, interfacial instabilities and patterns emerge in a wide variety of systems: phase-separated droplets driven by chemical reactions \cite{Zwicker2017,Seyboldt2018}, microswimmer suspensions \cite{Driscoll2017,Patteson2018,Miles2019}, chemotactic cells \cite{Bhattacharjee2021,Alert2021e}, growing tumors \cite{Greenspan1976,Khain2006,Basan2011a,Nagilla2018,Bogdan2018,Martin2021c} and bacterial biofilms \cite{Ben-Jacob2000,Allen2019,Kitsunezaki1997,Muller2002,Farrell2013,BenAmar2013,BenAmar2016a,Doostmohammadi2016a,Wang2017a,Trinschek2018,Yaman2019}, as well as in epithelial monolayers, either in mechanical competition \cite{Williamson2018,Buscher2020} or spreading freely \cite{Alert2020,Perez-Gonzalez2019,Alert2019,Trenado2021}. The instability mechanisms are varied; sometimes they rely on activity regulation, for example through nutrient depletion \cite{Greenspan1976,Wang2017a}, surfactant production \cite{Trinschek2018}, mechanical regulation of cell growth \cite{Williamson2018,Martin2021c}, and limitations in chemical sensing \cite{Alert2021e}. In other cases, the instability results just from the interplay of active forces and interface dynamics. For example, active polar forces can produce waves on the surface of fluid films and membranes \cite{Sankararaman2009,Sarkar2012,Sarkar2013,Maitra2014,Yang2014}, and they can destabilize the interface of a spreading tissue \cite{Alert2019}. Similarly, the free surface of an active nematic can also be destabilized by its bulk activity \cite{Blow2017,Alonso-Matilla2019,Soni2019,Liang2020,Lin2021,Thijssen2021c}, and many simulations showcased the unsteady shape dynamics of active nematic interfaces \cite{Mueller2021,Blow2014,Giomi2014,Doostmohammadi2016a,Fialho2017,Gao2017a,Metselaar2019,Coelho2019,Coelho2020,Ruske2021,Thijssen2021c}.

Crucially, the interface of active droples not only provides a free boundary but can also localize activity. Activity localization can happen either because the interface induces nematic order, as when cells align at the tissue boundary \cite{Doxzen2013,Duclos2017,Bade2018,Comelles2021,Xie2021,Zhang2021g} (\cref{Fig cell-monolayers}), or because the active material adsorbs at the interface, as is the case of the cell cortex and of synthetic active films in either vesicles or at oil-water interfaces \cite{Sanchez2012,Keber2014,Guillamat2018,Chen2021a}. Here, I take this localization effect into account by considering a fluid that is isotropic in the bulk but acquires nematic order at the interface. The order then decays inward over a penetration length. Tuning this penetration length with respect to droplet size allows the theory to encompass situations in which the order either remains interfacial or spans the entire droplet.

Similarly, I also account for an arbitrary anchoring angle, i.e., the angle between the local axis of nematic order and the interface. This is relevant as biological systems can exhibit a wide variety of anchoring angles. For example, microtubules tend to orient parallel to oil-water interfaces \cite{Chen2021a} (\cref{Fig active-vesicles}). Actin filaments typically orient roughly parallel to the membrane in the cell cortex but roughly perpendicular to the membrane in protrusions like the lamellipodium \cite{Blanchoin2014}. Similarly, the orientation of stress fibers is dynamical and coupled to cell shape \cite{Schakenraad2020}. At the multicellular scale, depending on the cell type and experimental conditions, epithelial and mesenchymal cells can align either parallel \cite{Doxzen2013,Duclos2017,Bade2018,Comelles2021,Xie2021} (\cref{Fig cell-monolayers}), perpendicular \cite{Perez-Gonzalez2019}, or at an intermediate angle \cite{Duclos2018} with respect to the tissue boundary. Cell monolayers can even dynamically reorganize from one boundary condition to another \cite{Guillamat2022}. Instead, bacteria tend to orient parallel to the biofilm edge \cite{Dellarciprete2018,Zhang2021g} (\cref{Fig bacterial-biofilms}), partly as a result of the active anchoring phenomenon \cite{Blow2014,Blow2017}. Finally, because of hydrodynamic torques, pusher and puller swimmers reorient differently at interfaces \cite{Huang2020a}. To account for this diversity of anchoring conditions, here we can tune the anchoring angle to interpolate between planar (parallel) and homeotropic (perpendicular) anchoring.

Altogether, I obtain results in terms of three dimensionless parameters: the nematic penetration length relative to droplet size, the anchoring angle, and the active capillary number, which compares active forces to surface tension. Depending on the anchoring angle and the extensile/contractile nature of the active stresses, droplets can break their initial circular symmetry above a critical active capillary number, which can be reached by increasing either activity or droplet size. Increasing activity localization, for example by decreasing the nematic penetration length, favors stability. The selected mode of the instability, which determines the number of fingers in the resulting pattern, increases with activity but varies non-monotonically with the nematic penetration length. The number of fingers increases as the nematic penetration increases while remaining small compared to droplet size. However, when the penetration length becomes comparable to the droplet size, the range of unstable modes shrinks, and modes with fewer fingers are selected again. Overall, this work provides a minimal analytical theory for the morphological stability of active nematic droplets, which exposes the underlying mechanisms, complements existing simulations, and can help interpret experimental observations.

\section{Model of an active nematic droplet} \label{model}

To analyze basic mechanisms of their morphological stability, I study a minimal model of active nematic droplets. I consider a two-dimensional circular droplet of incompressible fluid on a substrate.

\textbf{Nematic order.} The orientational degrees of freedom of the fluid are described in terms of the nematic order parameter tensor $\bm{Q}$. In two dimensions, $Q_{\alpha\beta} = S \,[ 2\,\hat{n}_\alpha \hat{n}_\beta - \delta_{\alpha\beta}]$, where $S$ is the scalar strength of the order parameter, and $\hat{\bm{n}} = (\cos\theta,\sin\theta)$ is the unitary director field, with $\theta$ the orientation angle \cite{deGennes-Prost}. The cartesian components of $\bm{Q}$ are $Q_{xx} = - Q_{yy} = S \cos(2\theta)$ and $Q_{xy} = Q_{yx} = S \sin(2\theta)$. In terms of $\bm{Q}$, and in the usual one-constant approximation of the Frank elastic energy, the nematic free energy reads \cite{Beris1994,Selinger2016}
\begin{equation} \label{eq free-energy}
F = \int \left[\frac{a}{2} Q_{\alpha\beta} Q_{\alpha\beta} + \frac{L}{2} (\partial_\alpha Q_{\beta\gamma})(\partial_\alpha Q_{\beta\gamma})\right]\,\dd^2\bm{r},
\end{equation}
where I take $a>0$ to stabilize the isotropic phase in the bulk, and $L$ is the orientational elastic modulus, which is directly related to the Frank elastic constant.

For simplicity, I ignore flow alignment of the nematic orientation. Moreover, I focus on flows over time scales longer than the nematic relaxation time $\gamma/a$, with $\gamma$ the rotational viscosity, so that the order parameter field rapidly relaxes to the equilibrium configuration given by $\delta F/\delta Q_{\alpha\beta}=0$. Given that $\bm{Q}$ is a rank-2 symmetric and traceless tensor, it can be described in terms of a single complex field $\chi = Q_{xx} + i Q_{xy} = S\, e^{2i\theta}$. In terms of this field, the equilibrium condition reads
\begin{equation} \label{eq chi}
\ell^2 \nabla^2 \chi = \chi,
\end{equation}
where I have defined the nematic length $\ell = \sqrt{L/a}$ that controls variations in nematic order through the droplet.

\begin{figure}[tbp]
\begin{center}
  {\phantomsubcaption\label{Fig planar-anchoring}}
  {\phantomsubcaption\label{Fig homeotropic-anchoring}}
\includegraphics[width=\columnwidth]{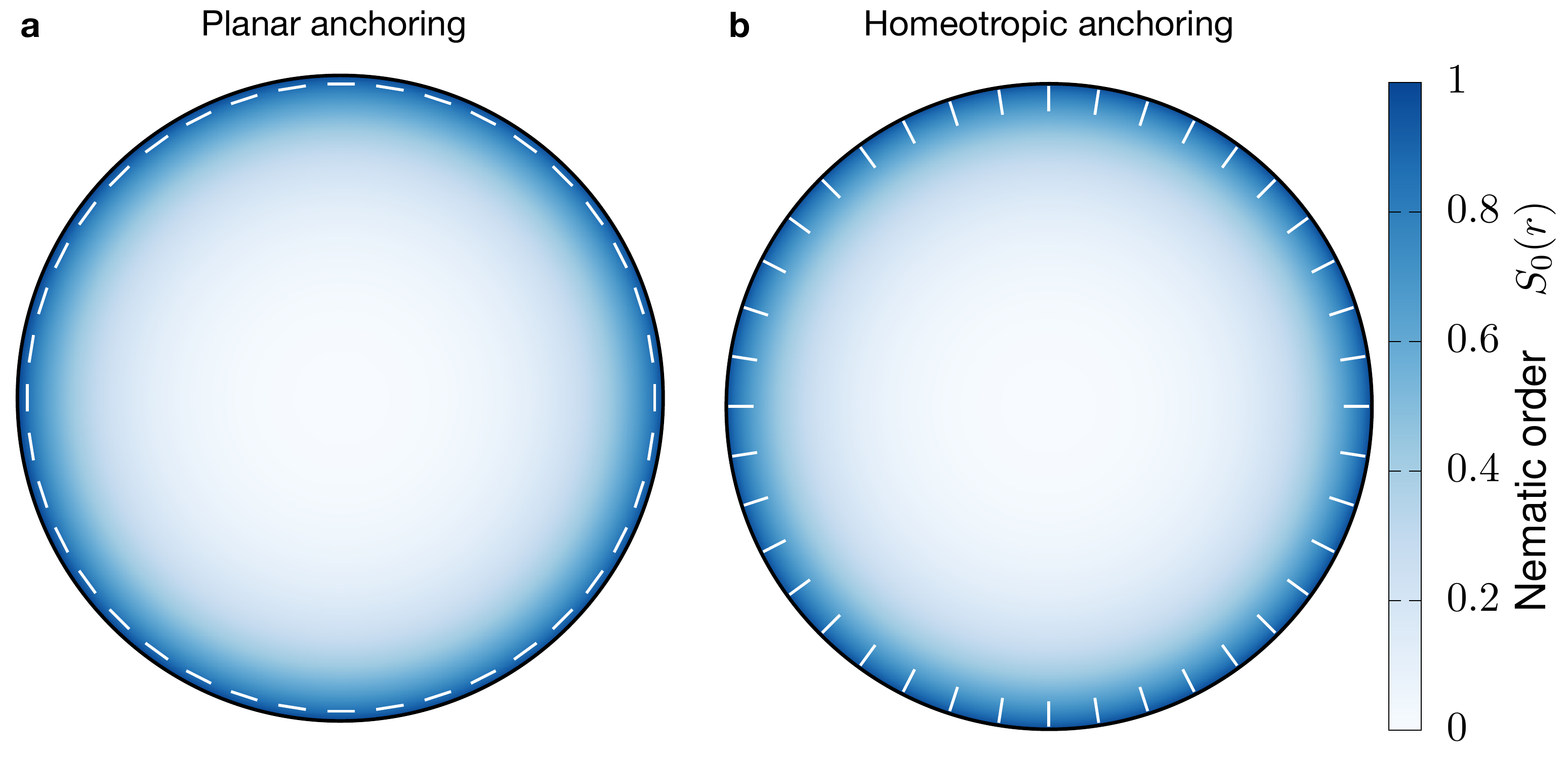}
\end{center}
\bfcaption{Nematic order in active droplets}{ The color map indicates the strength $S_0(r)$ of the nematic order in a droplet of radius $R_0 = 5\ell$. The white bars indicate the nematic angle $\theta(\phi)$ for planar (\subref*{Fig planar-anchoring}, $\theta_{\text{a}} = \pi/2$) and homeotropic (\subref*{Fig homeotropic-anchoring}, $\theta_{\text{a}} = 0$) anchoring.} \label{Fig unperturbed-droplets}
\end{figure}

As motivated in the introduction, I assume that the fluid acquires nematic order at the droplet interface. Specifically, I impose that the nematic order has maximal strength at the droplet boundary: $\left.S\right|_{r=R} = 1$. Solving \cref{eq chi}, I obtain (\cref{unperturbed-order})
\begin{equation} \label{eq S0}
S_0(r) = \frac{I_2(r/\ell)}{I_2(R_0/\ell)},
\end{equation}
where the subscript $0$ indicates that this is the reference solution for an unperturbed circular droplet of radius $R_0$. This solution shows that nematic order decays from the edge toward the center over a length $\ell$ to achieve the isotropic bulk state imposed by the free energy in \cref{eq free-energy} (\cref{Fig unperturbed-droplets}).

Respectively, the director $\hat{\bm{n}}$ anchors to the interface at an angle $\theta_{\text{a}}$ with respect to the normal vector $\hat{\bm{m}}$:
\begin{equation} \label{eq anchoring}
\left. \hat{\bm{n}}\cdot\hat{\bm{m}} \right|_{r=R} = \cos\theta_{\text{a}}.
\end{equation}
Thus, planar (parallel) anchoring corresponds to $\theta_{\text{a}} = \pi/2$, and homeotropic (perpendicular) anchoring corresponds to $\theta_{\text{a}} = 0$ (\cref{Fig unperturbed-droplets}). For an unperturbed circular droplet, the nematic angle throughout the droplet is independent of the radial coordinate and reads
\begin{equation} \label{eq theta0}
\theta_0(\phi) = \theta_{\text{a}} + \phi.
\end{equation}

\textbf{Force balance.} Active nematics generate an active anisotropic stress $\sigma_{\alpha\beta}^{\text{a}} = -\zeta Q_{\alpha\beta}$, with coefficient $\zeta>0$ for extensile and $\zeta<0$ for contractile active stresses. I ignore antisymmetric nematic stresses, which are of higher order in gradients with respect to active stresses. In the limit of fast nematic relaxation, as assumed to obtain \cref{eq chi}, flow-alignment stresses vanish. In the thin-film limit, viscous stresses are dominated by velocity gradients perpendicular to the film, which lead to a Darcy friction term $\xi \bm{v}$ when the flow is averaged over the film height. Altogether, force balance reduces to
\begin{equation} \label{eq force-balance}
-\bm\nabla P + \bm f^{\text{a}} = \xi \bm v,
\end{equation}
where $f^{\text{a}}_\alpha = -\partial_\beta \sigma^{\text{a}}_{\alpha\beta}$ is the active force density arising from gradients of the active stress, and $P$ is the pressure field that enforces the incompressibility condition $\bm\nabla \cdot \bm v = 0$. To leverage this condition, I take the divergence of \cref{eq force-balance} and obtain
\begin{equation} \label{eq pressure}
\nabla^2 P = \bm \nabla \cdot \bm f^{\text{a}} \equiv s.
\end{equation}
\Cref{eq pressure} is a Poisson equation for the pressure field, where the divergence of the active force density acts as a pressure source $s$.

At the droplet interface, with normal vector $\hat{\bm m}$, I impose a line tension $\gamma$, which gives a discontinuity of the normal stress as prescribed by the Young-Laplace law:
\begin{equation} \label{eq Young-Laplace}
\left. \hat{m}_\alpha\, \sigma_{\alpha\beta}\, \hat{m}_\beta \right|_{r=R} = - \gamma \left.\bm{\nabla}\cdot \hat{\bm{m}}\right|_{r=R}.
\end{equation}
Here, $\sigma_{\alpha\beta} = -P\delta_{\alpha\beta} -\zeta Q_{\alpha\beta}$ is the total stress tensor of the active fluid. I have assumed that the external fluid is ideal, and I have set the pressure origin so that $P_{\text{ext}}=0$. Solving \cref{eq pressure} with these conditions, I obtain a pressure profile (\cref{unperturbed-flow})
\begin{equation} \label{eq P0}
P_0(r) = \frac{\gamma}{R_0} - \zeta \cos(2\theta_{\text{a}}) \left[1 - \frac{I_0(R_0/\ell) - I_0(r/\ell)}{I_2(R_0/\ell)}\right].
\end{equation}
Introducing this solution into the force balance \cref{eq force-balance}, I obtain a vanishing velocity $v_r^0(r) = 0$. The unperturbed droplet is quiescent: The pressure gradient and the line tension exactly compensate the active force without inducing flow.

\begin{figure}[tbp]
\begin{center}
\includegraphics[width=\columnwidth]{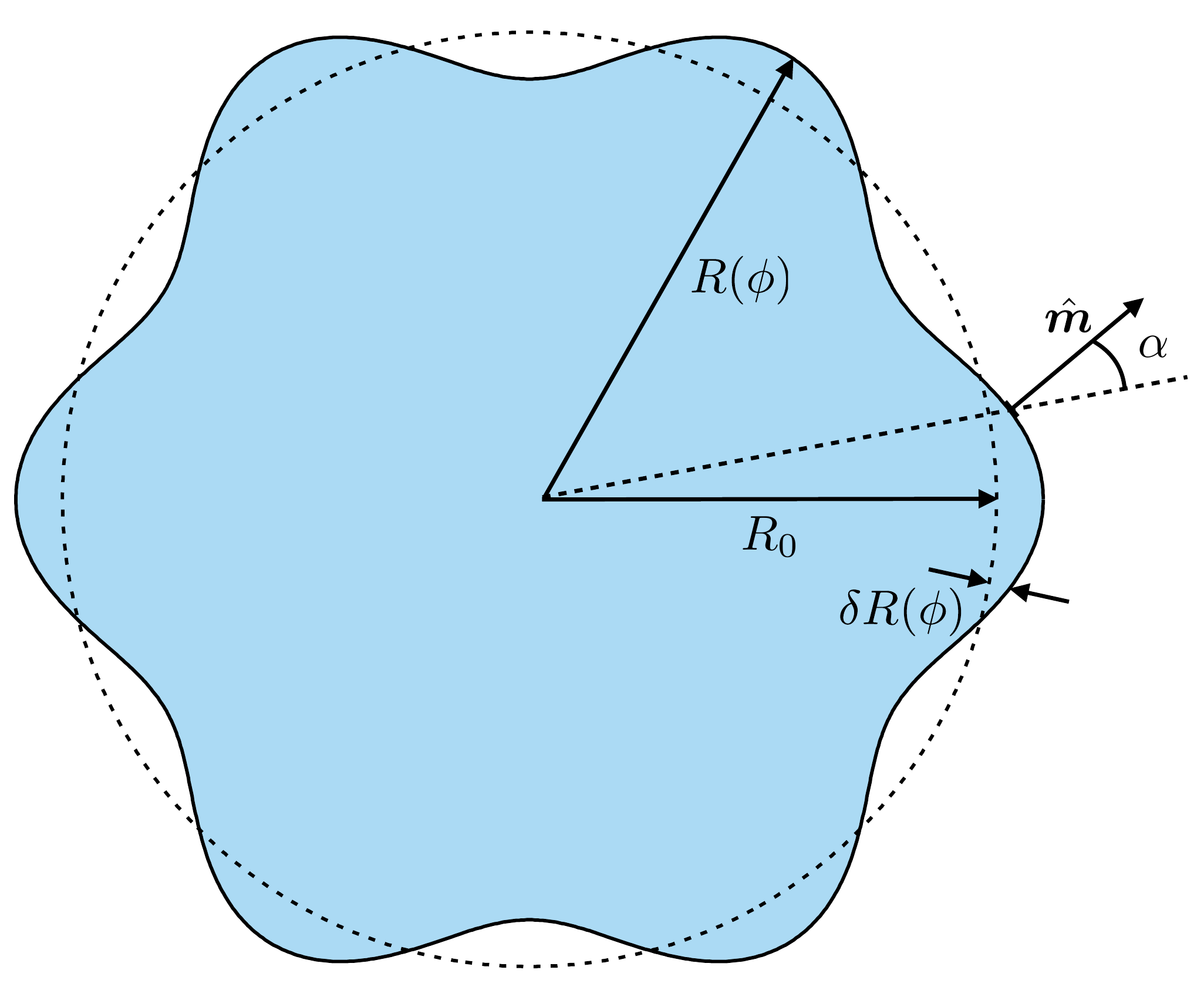}
\end{center}
\bfcaption{Shape perturbations of a circular droplet}{ The dashed circle indicates the unperturbed shape.} \label{Fig perturbations}
\end{figure}

\section{Morphological stability} \label{stability}

\textbf{Shape perturbations and growth rate.} To analyze the linear stability of the circular droplet shape, I introduce morphological perturbations by allowing the droplet radius to vary with the polar angle (\cref{Fig perturbations}): $R(\phi) = R_0 + \delta R(\phi)$. Accordingly, the strength and orientation of the nematic order are also perturbed (\cref{perturbed-order}), as are the forces and flows (\cref{perturbed-flow}). The flow induced by the perturbations then drives interface motion through the kinematic condition
\begin{equation} \label{eq kinematic-condition}
\frac{\dd R(\phi)}{\dd t} = \left.\bm{v}\cdot \hat{\bm{m}}\right|_{r=R} \approx \delta v_r(R_0,\phi),
\end{equation}
where $\hat{\bm{m}}$ is the normal vector, and I have expanded to first order in perturbations and used that $v_r^0(r) = v_\phi^0(r) = 0$. To analyze the interface dynamics, I decompose all fields in angular Fourier modes labelled by the index $k$, which indicates the number of protrusions of the perturbed droplet contour (\cref{Fig perturbations}). The growth rate of the morphological perturbations is given by $\omega_k = \delta\tilde{v}_{r,k}(R_0)/\delta\tilde{R}_k$. Introducing all the perturbation results obtained in \cref{perturbed}, I obtain the final result for the growth rate, whose complete expression is given in \cref{interface-dynamics}.

\begin{figure}[tbp]
\begin{center}
\includegraphics[width=\columnwidth]{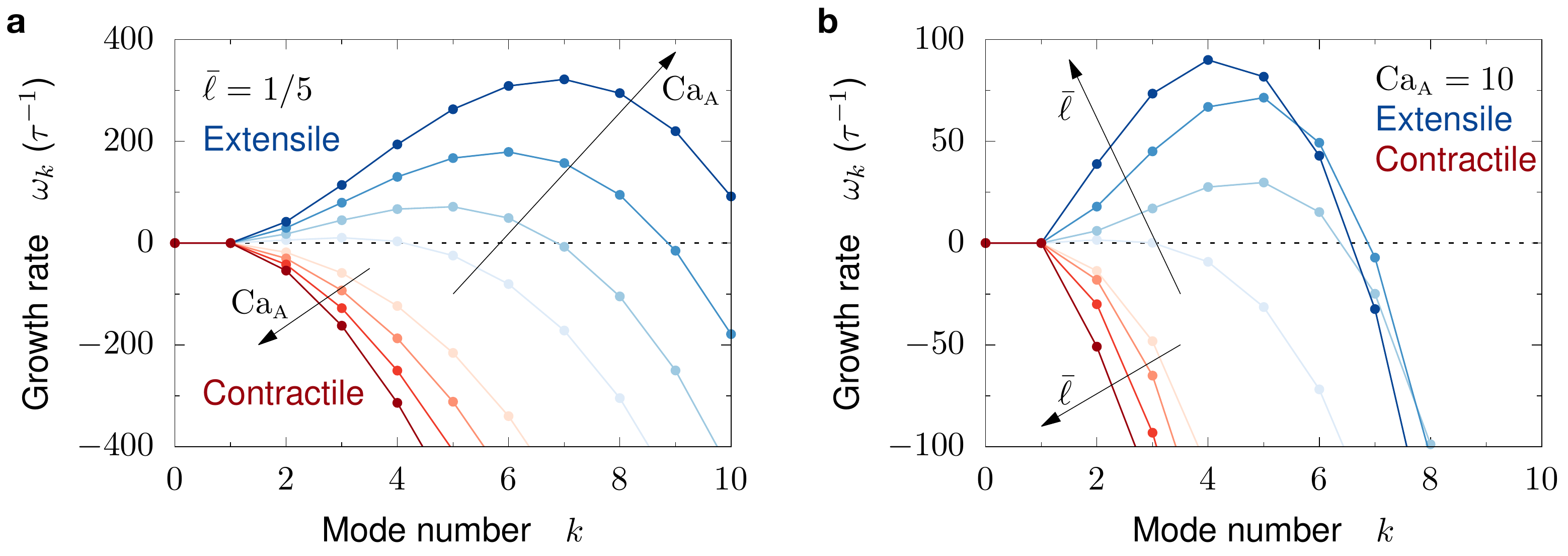}
\end{center}
  {\phantomsubcaption\label{Fig growth-rate-capillary}}
  {\phantomsubcaption\label{Fig growth-rate-ell}}
\bfcaption{Growth rate of shape perturbations of active nematic droplets}{ When rescaled by the capillary time $\tau \equiv \xi R_0^3/\gamma$, the growth rate depends only on three dimensionless parameter combinations: the extensile or contractile sign of active stresses, the rescaled nematic length $\bar\ell \equiv \ell/R_0$, and the active capillary number $\mathrm{Ca}_{\text{A}}\equiv |\zeta|R_0/\gamma$. These plots are for planar anchoring (\cref{Fig planar-anchoring}). For homeotropic anchoring (\cref{Fig homeotropic-anchoring}), the results are equivalent upon exchanging extensile for contractile. \subref*{Fig growth-rate-capillary}, For extensile (contractile) active stresses, increasing the active capillary number leads to a further destabilization (stabilization) of the droplet shape. The values of the active capillary number are $\mathrm{Ca}_{\text{A}} = 5n$; $n=1,\ldots,4$. \subref*{Fig growth-rate-ell}, A similar trend holds as the rescaled nematic length increases. Its values are $\bar\ell = 2^n/20$; $n=0,\ldots,3$.} \label{Fig growth-rate}
\end{figure}

The result shows that the stability of droplets with planar anchoring and extensile stresses is equivalent to that of droplets with homeotropic anchoring and contractile stresses. More generally, the growth rate is invariant under the transformation $\zeta, \theta_{\text{a}} \rightarrow -\zeta,\theta_{\text{a}} + \pi/2$. For clarity, hereafter I discuss the results for planar anchoring $\theta_{\text{a}} = \pi/2$. In this case, the growth rate simplifies to
\begin{multline} \label{eq growth-rate}
\omega_k = \frac{k(k-1)}{\xi R_0^3} \left\{ -\gamma (k+1) \phantom{+ 2\zeta R_0 \left[\frac{2(k-1)\ell}{R_0} + \frac{I_1(R_0/\ell)}{I_2(R_0/\ell)}\right]} \right.\\
\left.+ 2\zeta R_0 \left\{ \left[\frac{2(k-1)\ell}{R_0} + \frac{I_1(R_0/\ell)}{I_2(R_0/\ell)}\right]\frac{I_{k-1}(R_0/\ell)}{I_{k-2}(R_0/\ell)} - 1\right\} \right\}.
\end{multline}
The prefactor ensures that both the dilation/contraction mode $k=0$ and the translation mode $k=1$ are marginal, $\omega_0 = \omega_1 = 0$, as a consequence of the fluid's incompressibility and translational invariance, respectively. Beyond the prefactor, the first term in \cref{eq growth-rate} corresponds to the stabilizing contribution of line tension, whereas the second term accounts for the active effects, which are stabilizing (destabilizing) for contractile (extensile) stresses (\cref{Fig growth-rate}).

The growth rate \cref{eq growth-rate} depends on four parameter combinations: a capillary time $\tau \equiv \xi R_0^3/\gamma$, the ratio of nematic length and droplet size $\bar\ell\equiv \ell/R_0$, the sign of the active stress $\zeta/|\zeta|$, and the active capillary number $\mathrm{Ca}_{\text{A}} \equiv |\zeta| R_0/\gamma$. Using the capillary time as the time unit, the rescaled growth rate $\bar\omega_k\equiv \omega_k\tau$ can be recast in terms of the other three (dimensionless) parameters:
\begin{multline} \label{eq dimensionless-growth-rate}
\bar\omega_k = k(k-1)\left\{ - (k+1) \phantom{\frac{I_{k-1}(1/\bar\ell)}{I_{k-2}(1/\bar\ell)}} \right.\\
\left. + 2 \, \mathrm{Ca}_{\text{A}}\, \frac{\zeta}{|\zeta|} \left\{ \left[2(k-1) \bar\ell + \frac{I_1(1/\bar\ell)}{I_2(1/\bar\ell)}\right]\frac{I_{k-1}(1/\bar\ell)}{I_{k-2}(1/\bar\ell)} - 1\right\}\right\}.
\end{multline}

\textbf{Active capillary number.} The active capillary number compares active stresses to surface tension \cite{Giomi2014,Blow2017,Fialho2017,Alonso-Matilla2019}. It is an active variant of the ordinary capillary number $\mathrm{Ca} \equiv \eta V/\gamma$, which compares dissipative viscous forces to surface tension \cite{Guyon2001}. Here, $\eta$ is the shear viscosity, $V$ is a characteristic flow velocity, and $\gamma$ is the surface tension. In the present work, dissipation is due to friction, and hence the capillary number is instead defined as $\mathrm{Ca} \equiv \xi V L^2/\gamma$, where $L$ is a characteristic length of the droplet. Introducing the characteristic velocity of active flows, $V_{\text{A}}=|\zeta|/(\xi R_0)$, yields the active capillary number used here: $\mathrm{Ca}_{\text{A}} \equiv |\zeta| R_0/\gamma$.

Alternatively, this quantity can also be thought of as an active Bond number. The ordinary Bond number, also known as the E\"{o}tv\"{o}s number, compares graviational to surface tension forces, and it is used to characterize the shape of drops, for example during gravity-driven wetting \cite{Guyon2001}. It is defined as $\mathrm{Bo} \equiv \Delta \rho \,g L^2/\gamma$, where $\Delta \rho$ is the density difference between two media (e.g., the liquid of the droplet and the surrounding fluid), and $g$ is the gravitational acceleration. Unlike the ordinary capillary number, the Bond number compares a driving force (gravity) to surface tension, parallel to how the active capillary number compares active driving forces to surface tension. Furthermore, the active capillary number can be written as $\mathrm{Ca}_{\text{A}} = R_0/\ell_{\text{ac}}$, where $\ell_{\text{ac}} \equiv \gamma/|\zeta|$ is an active capillary length. This length is an active variant of the ordinary capillary length $\ell_{\text{c}} \equiv \sqrt{\gamma/(\Delta\rho\, g)}$ defined by the balance of gravitational and surface tension forces, which allows to write the Bond number as $\mathrm{Bo} = L^2/\ell_{\text{c}}^2$. Overall, the active capillary number, which here controls droplet shape stability, has conceptual parallels with both the capillary and Bond numbers of passive fluids.

\textbf{Stability diagram and mode selection.} How do these dimensionless parameters control droplet stability? For planar anchoring, \cref{Fig growth-rate} shows that droplets with contractile stresses ($\zeta <0$) are stable ($\omega_k <0$), whereas droplets with extensile stresses ($\zeta>0$) experience a morphological instability (some modes $k$ with $\omega_k >0$). The competition between active forces and line tension, controlled by both $\mathrm{Ca}_{\text{A}}$ and $\bar\ell$, governs the range of unstable modes and selects the mode with the fastest growth rate, which determines the initial number of fingers resulting from the instability.

For contractile stresses, both increasing the active capillary number (\cref{Fig growth-rate-capillary}) and increasing the rescaled nematic length (\cref{Fig growth-rate-ell}) result in a further stabilization of the droplet shape. In contrast, for extensile stresses, as the active capillary number $\mathrm{Ca}_{\text{A}}$ increases, more modes become unstable, and the selected mode becomes higher (\cref{Fig growth-rate-capillary}). The same trend is obtained when increasing the rescaled nematic length $\bar\ell$ while keeping it small, $\bar\ell \ll 1$. However, when the nematic length becomes comparable to the droplet radius, this behavior changes. As the rescaled nematic length is increased further, the range of unstable modes shrinks a bit, and the selected mode becomes lower again (\cref{Fig growth-rate-ell}). Eventually, in the limit of large nematic length $\bar\ell\rightarrow \infty$, which corresponds to nematic order extending throughout the droplet, the growth rate becomes independent of $\bar\ell$:
\begin{equation}
\lim_{\bar\ell\rightarrow \infty} \bar\omega_k = -k(k^2-1) + 4\, \mathrm{Ca}_{\text{A}}\, \frac{\zeta}{|\zeta|} \,k.
\end{equation}

\begin{figure}[tbp]
\begin{center}
\includegraphics[width=0.75\columnwidth]{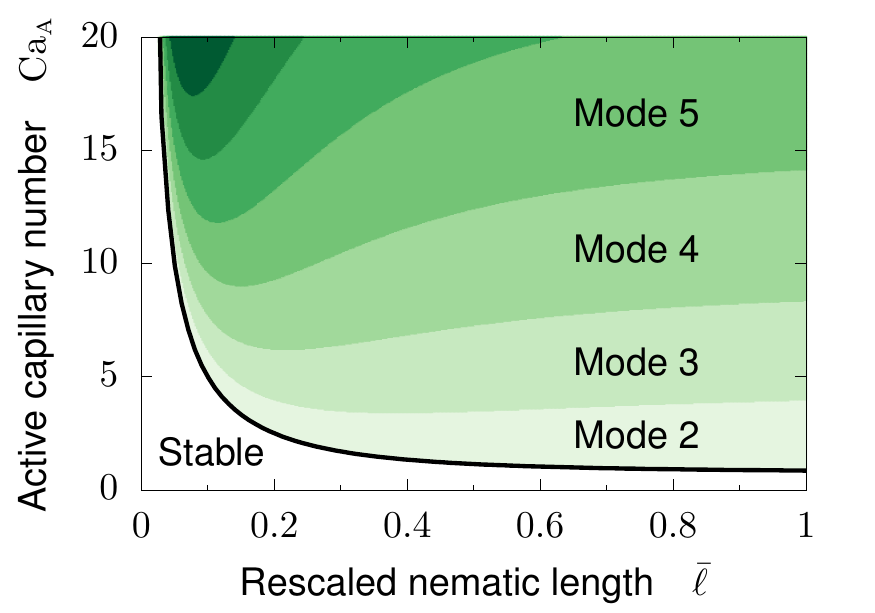}
\end{center}
\bfcaption{Morphological stability diagram of active nematic droplets}{ Below the critical active capillary number (\cref{eq critical-Ca}), a circular active nematic droplet is stable. Above the critical value, the droplet experiences a fingering instability. The selected (fastest-growing) mode increases monotonically with the active capillary number, but it exhibits a non-monotonic dependence on the nematic length. This plot is for either extensile stresses and planar anchoring (\cref{Fig planar-anchoring}), or equivalently contractile stresses and homeotropic anchoring (\cref{Fig homeotropic-anchoring}). In the remaining combinations, the droplet is stable. In cases with intermediate anchoring angles, the results follow from the expressions in \cref{interface-dynamics}.} \label{Fig stability-diagram}
\end{figure}

As seen in either \cref{eq growth-rate} or \cref{eq dimensionless-growth-rate}, the destabilizing active effects dominate at long wavelengths (low mode number $k$). Hence, an infinite interface would always be unstable. However, for the interface of a finite droplet, the first mode that might become unstable is the elliptic mode $k=2$. Therefore, for a droplet, the instability has a finite threshold, given by $\bar\omega_2 = 0$. The critical value of the active capillary number is
\begin{equation} \label{eq critical-Ca}
\mathrm{Ca}_{\text{A}}^{\text{c}} = \frac{3}{2}\frac{I_0(1/\bar\ell) I_2(1/\bar\ell)}{I_1^2(1/\bar\ell) - I_2^2(1/\bar\ell)},
\end{equation}
which monotonically decreases with the rescaled nematic length (black curve in \cref{Fig stability-diagram}). Instability is most favorable in the limit of large nematic length compared to droplet size, in which the critical active capillary number tends to its minimum: $\lim_{\bar\ell\rightarrow \infty} \mathrm{Ca}_{\text{A}}^{\text{c}} = 3/4$. In the unstable region, the stability diagram in \cref{Fig stability-diagram} also shows the selected mode. As explained earlier, the selected mode increases monotonically with the active capillary number, but it features a non-monotonic behavior with the nematic length $\ell$. Respectively, increasing droplet size $R_0$ decreases $\bar\ell = \ell/R_0$ but increases $\mathrm{Ca}_{\text{A}} = |\zeta| R_0/\gamma$, and hence it corresponds to moving up along a hyperbola in the stability diagram in \cref{Fig stability-diagram}. Thus, the selected mode also varies non-monotonically with droplet size.

\section{Discussion and outlook} \label{discussion}

\begin{figure}[tbp]
\begin{center}
\includegraphics[width=0.75\columnwidth]{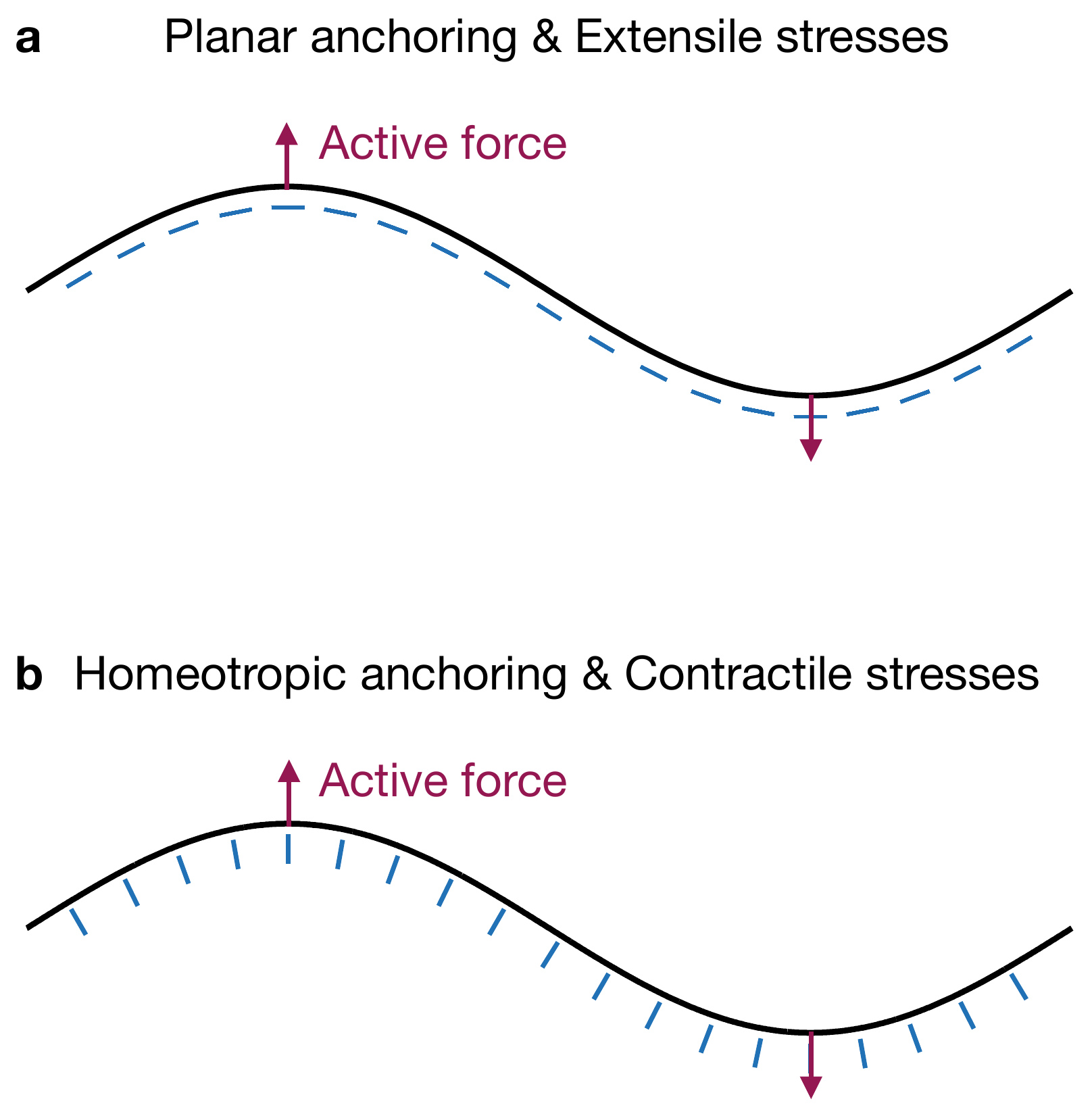}
\end{center}
  {\phantomsubcaption\label{Fig planar-extensile}}
  {\phantomsubcaption\label{Fig homeotropic-contractile}}
\bfcaption{Instability mechanism}{ Schematic of the interface (black), the underlying director field (blue), and the active forces that it generates on the interface (red). Certain combinations of anchoring conditions and active stresses amplify shape perturbations, leading to droplet shape instability. The figure depicts the unstable cases of \subref*{Fig planar-extensile}, planar anchoring ($\theta_{\text{a}} = \pi/2$) and extensile stresses ($\zeta>0$), and \subref*{Fig homeotropic-contractile}, homeotropic anchoring ($\theta_{\text{a}}=0$) and contractile stresses ($\zeta<0$).} \label{Fig mechanism}
\end{figure}

I have shown that active nematic droplets can experience a morphological instability. The mechanism is simple: Droplet shape perturbations distort the nematic order, which generates an active force that further deforms the droplet (\cref{Fig mechanism}). This mechanism is similar to that of the well-known bulk instability of active nematics, which leads to spontaneous flows even in unbounded systems without an interface \cite{Ramaswamy2010,Marchetti2013,Alert2022}. Despite the similarities, the instability presented here is interfacial in nature, and thus it is fundamentally different from the bulk instability.

In the bulk instability, a perturbation in the nematic director generates active flows that further rotate the director. This feedback requires the director field to have a dynamics, which couples it directly to the flow. Here, instead, I have taken the nematic order to instantaneously relax to its equilibrium configuration (\cref{eq chi}), and hence the director has no intrinsic dynamics \cite{Lin2021}. Therefore, the active nematic considered here cannot experience a bulk instability. Yet, it can be unstable in the presence of an interface. Through anchoring, interface motion affects the director field, and therefore it provides the missing dynamical field that enables the feedback between the director and active flows. Previous works considered similar interfacial instabilities but retained one additional dynamical field, either the concentration of microswimmers \cite{Alonso-Matilla2019}, the director \cite{Soni2019}, or the density in a compressible active nematic \cite{Lin2021}. Thus, the theory presented here provides a minimal description of morphological instability in active nematics, in which interface motion is the only dynamical field.

\pagebreak

The instability takes place only for appropriate combinations of the anchoring angle and the sign of the active stresses, as illustrated in \cref{Fig mechanism}. For these combinations, active stresses tend to extend the interface, which consequently undulates and forms finger-like protrusions. Beyond the initial, linear stage of the instability, non-linear effects arising from line tension and incompressibility could potentially saturate finger growth. In this case, an active instability would lead to a static fingering pattern, similar to the recently-found buckling instability in active nematic films \cite{Senoussi2019b}, but in stark contrast to the flowing steady states characteristic of active matter. Alternatively, the fingering process could lead to pinch-off events and droplet splitting \cite{Giomi2014}.

The fingering instability presented here has implications for both biological and synthetic active systems. Foremost, it provides a symmetry-breaking mechanism for the spontaneous shape changes observed in multiple systems, from the sub-cellular scales of the mitotic spindle and the cell cortex to the scale of entire organisms such as \textit{Hydra}, and including reconstituted systems such as active vesicles (\cref{Fig1}). The findings might be particularly relevant for epithelial cell monolayers. In situations such as wound healing, cells at the tissue edge polarize perpendicularly to the interface, which creates active fingering instabilities \cite{Alert2019}. In other situations, however, cells align parallel to the interface. The results of this work could help understand the tissue shape changes observed in these cases \cite{Comelles2021}. The reported fingering instability could also be exploited to pattern active materials, for example to design corrugated surfaces with potential applications as reconfigurable substrates to study tissue dynamics.

Looking forward, this work could be extended to capture the three-dimensional profile of the droplet \cite{Joanny2012}, which would bring in additional effects such as wetting energies \cite{Trinschek2017,Trinschek2020a} and out-of-plane nematic order \cite{Nejad2022,Hoffmann2021}. Other interesting extensions would be to include chiral flows \cite{Soni2019a}, mechanochemical processes \cite{Mietke2019b,Mietke2019a}, and an external elastic medium, which is relevant to study the growth of biofilms in mucus-like gels and in host tissues \cite{Zhang2021g}.

\bigskip

\textbf{Acknowledgments.} I thank Jaume Casademunt, John D. McEnany, Howard A. Stone, Ned S. Wingreen, and Jing Yan for discussions.

\bibliography{Active_droplets}

\onecolumngrid 

\clearpage
\appendix



\twocolumngrid

\label{SI}

\section{Unperturbed state. Circular droplet} \label{unperturbed}

As a reference, I consider a circular droplet of radius $R_0$.

\subsection{Nematic order} \label{unperturbed-order}

In polar coordinates, the equilibrium condition \cref{eq chi} for the nematic order reads
\begin{equation} \label{eq chi-polar}
\left[\partial_r^2 + \frac{1}{r}\partial_r + \frac{1}{r^2}\partial_\phi^2 - \frac{1}{\ell^2}\right] \chi(r,\phi) = 0.
\end{equation}
The anchoring condition imposes the nematic angle at the droplet boundary: $\theta(R_0,\phi)= \theta_{\text{a}} + \phi$, where $\phi$ is the polar angle. Because there is no reason for the nematic angle to change along the radial coordinate, the nematic angle is a function of the polar angle only, as given in \cref{eq theta0}. Similarly, axial symmetry implies that the strength of the nematic order depends only on the radial coordinate, $S=S_0(r)$. Therefore, $\chi(r,\phi)=S_0(r) \,e^{i2\theta_0(\phi)}$, and \cref{eq chi-polar} reduces to
\begin{equation}
S''_0(r) +\frac{1}{r}S'_0(r) - \left[\frac{1}{\ell^2} + \frac{4}{r^2}\right]S_0(r) = 0.
\end{equation}
The solutions to this equation are modified Bessel functions of order $n=2$ and scale factor $\ell$. Imposing that the nematic order has a maximal strength $S_0(R_0)=1$ at the boundary, I obtain the solution in \cref{eq S0}, which is displayed in \cref{Fig unperturbed-droplets}.

\subsection{Forces and flows} \label{unperturbed-flow}

For the nematic order of the unperturbed droplet, given by \cref{eq theta0,eq S0}, the active force density has components
\begin{subequations}
\begin{align}
f_r^{\text{a},0}(r) &= -\zeta \cos(2\theta_{\text{a}}) \left[ S'_0(r) +\frac{2}{r} S_0(r)\right],\\
f_\phi^{\text{a},0}(r) &= 0.
\end{align}
\end{subequations}
Given that the active force has axial symmetry, the pressure is also axially symmetric, $P_0(r,\phi) = P_0(r)$. Hence, the Poisson equation for the pressure field reads as
\begin{equation} \label{eq Poisson-0}
\left[\frac{\dd^2}{\dd r^2} + \frac{1}{r}\frac{\dd}{\dd r}\right] P_0(r) = -\zeta \cos(2\theta_{\text{a}}) \left[ S''_0(r) + \frac{3}{r} S'_0(r)\right].
\end{equation}
To solve this equation, we need to impose boundary conditions as specified by the Young-Laplace relation in \cref{eq Young-Laplace}.

The stress tensor of the active fluid is given by
\begin{equation}
\sigma_{\alpha\beta} = -P\delta_{\alpha\beta} -\zeta Q_{\alpha\beta},
\end{equation}
and its components in a cylindrical-coordinate basis are
\begin{subequations}
\begin{align}
\sigma_{rr} &= -P -\zeta S \cos(2(\theta - \phi)),\\
\sigma_{r\phi} & = \sigma_{\phi r} = -\zeta S \sin(2(\theta - \phi)),\\
\sigma_{\phi\phi}& = -P +\zeta S \cos(2(\theta-\phi)).
\end{align}
\end{subequations}
For the unperturbed circular droplet, $\sigma^0_{rr}(r) = -P_0(r) - \zeta \cos(2\theta_{\text{a}}) S_0(r)$, and hence, the Young-Laplace boundary condition \cref{eq Young-Laplace} becomes
\begin{equation}
P_0(R_0) = \frac{\gamma}{R_0} - \zeta \cos(2\theta_{\text{a}}).
\end{equation}
With this boundary condition, the solution to \cref{eq Poisson-0} is given in \cref{eq P0}.

\section{Pertubed state. Non-circular droplet} \label{perturbed}

Here, I obtain the perturbations in nematic order, forces, and flows induced by perturbations in droplet shape, which are introduced as a radius that varies along the droplet contour, i.e. with the polar angle $\phi$: $R(\phi) = R_0 + \delta R(\phi)$.

\subsection{Nematic order} \label{perturbed-order}

As a result of the morphological perturbations, the strength and orientation of the nematic order are perturbed as
\begin{subequations}
\begin{align}
S(r,\phi) &= S_0(r) + \delta S(r,\phi),\\
\theta(r,\phi) &= \theta_0(\phi) + \delta\theta(r,\phi).
\end{align}
\end{subequations}
Hence, to first order in the perturbations, the complex field $\chi$ reads
\begin{multline} \label{eq chi-expansion}
\chi(r,\phi) = S(r,\phi) e^{i2\theta(r,\phi)} \\
\approx \left[S_0(r) + \delta S(r,\phi) + 2 i S_0(r) \,\delta\theta(r,\phi)\right]\,e^{i2\theta_0(\phi)}.
\end{multline}
Using this expression, the nematic equilibrium condition \cref{eq chi-polar} becomes a complex equation for the perturbation fields $\delta S$ and $\delta\theta$. The real and imaginary parts of this equation must vanish separately, which leads to the following pair of coupled partial differential equations (PDEs):
\begin{subequations}
\begin{gather}
\begin{multlined}
\left[ \partial_r^2 + \frac{1}{r}\partial_r + \frac{1}{r^2} \left[\partial^2_\phi -4\right] - \frac{1}{\ell^2}\right] \delta S(r,\phi) \\
- \frac{8}{r^2} S_0(r)\,\partial_\phi \delta\theta(r,\phi) = 0,
\end{multlined}
\\
\begin{multlined}
\frac{2}{r^2}\partial_\phi \delta S(r,\phi) + \left[S_0(r) \,\partial_r^2 + \left[2 S'_0(r) +\frac{1}{r} S_0(r)\right] \partial_r \right.\\
\left. + \frac{1}{r^2} S_0(r)\, \partial^2_\phi\right] \delta\theta(r,\phi) = 0.
\end{multlined}
\end{gather}
\end{subequations}

To solve these equations, we introduce the angular Fourier decomposition of the perturbation fields:
\begin{subequations}
\begin{align}
\delta S(r,\phi) &= \sum_{k=0}^\infty \delta\tilde{S}_k (r)\,e^{ik\phi},\\
\delta \theta(r,\phi) &= \sum_{k=0}^\infty \delta\tilde{\theta}_k (r)\,e^{ik\phi}.
\end{align}
\end{subequations}
In terms of their Fourier components, the pair of coupled PDEs becomes a pair of coupled ordinary differential equations (ODEs):
\begin{subequations} \label{eq perturbations-Fourier}
\begin{gather}
\begin{multlined}
\left[\frac{\dd^2}{\dd r^2} +\frac{1}{r}\frac{\dd}{\dd r} - \frac{k^2 + 4}{r^2} - \frac{1}{\ell^2}\right] \delta \tilde{S}_k(r) \\
- \frac{8ik}{r^2} S_0(r)\,\delta\tilde{\theta}_k(r) = 0, \label{eq perturbations-real}
\end{multlined}
\\
\begin{multlined}
\frac{2ik}{r^2}\delta\tilde{S}_k(r) + \left[S_0(r)\frac{\dd^2}{\dd r^2} + \left[ 2 S'_0(r) +\frac{1}{r} S_0(r)\right]\frac{\dd}{\dd r} \right.\\
\left. - \frac{k^2}{r^2} S_0(r)\right]\delta\tilde{\theta}_k(r) = 0 \label{eq perturbations-imaginary}.
\end{multlined}
\end{gather}
\end{subequations}
Even though the coefficients of the differential operators involve non-linear functions, these equations can be solved analytically. The solution can be guessed by looking at the operator on $\delta\tilde{S}_k(r)$ in \cref{eq perturbations-real}. Adding an appropriate term, this operator would correspond to a modified Bessel operator of integer order. There are two symmetric ways to complete this operator to this end: either to add or to subtract a term $4k/r^2 \,\delta\tilde{S}_k(r)$, which respectively transform the operator into a modified Bessel operator of order $k\mp2$. Hence, I propose the ansatz
\begin{equation} \label{eq perturbed-S-ansatz}
\delta\tilde{S}_k(r) = A I_{k+2}(r/\ell) + B I_{k-2}(r/\ell).
\end{equation}
The corresponding modified Bessel functions of second species are also solutions of the equations. However, because they diverge at $r=0$, their integration constants must be set to zero. Introducing the ansatz into \cref{eq perturbations-real} yields an algebraic equation for $\delta\tilde{\theta}_k(r)$, whose solution is
\begin{equation} \label{eq perturbed-theta-ansatz}
\delta\tilde{\theta}_k(r) = \frac{i}{2}\frac{I_2(R_0/\ell)}{I_2(r/\ell)}\left[B I_{k-2}(r/\ell) - A I_{k+2}(r/\ell)\right].
\end{equation}
Although they were proposed for \cref{eq perturbations-real}, the solutions in \cref{eq perturbed-S-ansatz,eq perturbed-theta-ansatz} turn out to satisfy \cref{eq perturbations-imaginary}. Therefore, they are solutions to the full \cref{eq perturbations-Fourier}.

To determine the integration constants $A$ and $B$, we have to use boundary conditions. To this end, I derive the boundary conditions on the perturbed nematic order. First, I impose the anchoring condition, which depends on the normal vector $\hat{\bm{m}}$ of the perturbed boundary (\cref{Fig perturbations}). In terms of the local angle $\alpha$ between the perturbed and the original (circular) boundary, the normal vector reads
\begin{equation}
\hat{\bm{m}} = \cos\alpha\, \hat{\bm{r}} + \sin\alpha\, \hat{\bm{\phi}} \approx \hat{\bm{r}} - \frac{1}{R_0} \frac{\dd \delta R}{\dd \phi}\hat{\bm{\phi}}.
\end{equation}
Here, I have approximated $\hat{\bm{m}}$ to first order in perturbations using that $\alpha\approx \tan \alpha = -\dd \delta R/\dd s$, where $s=R_0\phi$ is the arc length coordinate. Then, in terms of the nematic director field $\hat{\bm{n}}$, the anchoring condition reads
\begin{equation} \label{eq n-boundary}
\hat{\bm{n}}(R,\phi)\cdot\hat{\bm{m}} = \cos \theta_{\text{a}}.
\end{equation}
Here, $\hat{\bm{n}}(r,\phi) = \hat{\bm{n}}_0(\phi) + \delta\hat{\bm{n}}(r,\phi)$
, where $\hat{\bm{n}}_0(\phi) = \cos\theta_{\text{a}}\,\hat{\bm{r}} + \sin\theta_{\text{a}}\,\hat{\bm{\phi}}$ and $\delta\hat{\bm{n}} = - \sin\theta_{\text{a}}\, \delta\theta\,\hat{\bm{r}} + \sin(\theta_{\text{a}} + 2\phi)\,\delta\theta\,\hat{\bm{\phi}}$. Thus, to first order in perturbations, \cref{eq n-boundary} implies
\begin{equation} \label{eq bc-theta}
\delta\theta(R_0,\phi) \approx -\frac{1}{R_0}\frac{\dd \delta R}{\dd\phi},
\end{equation}
which provides a boundary condition for the angle perturbations.

Next, I enforce that the strength of the nematic order remains 1 at the boundary, $S(R,\phi)=1$, which implies
\begin{equation}
S_0(R) + \delta S(R,\phi) =1.
\end{equation}
Taking into account that $S_0(R)\approx S_0(R_0) + S'_0(R_0)\delta R(\phi)$, and to first order in perturbations, I obtain the boundary condition for the nematic strength perturbations,
\begin{equation} \label{eq bc-S}
\delta S(R_0,\phi) \approx -S'_0(R_0)\delta R(\phi),
\end{equation}
where I have used that $S_0(R_0)=1$. The right-hand side can be evaluated using that
\begin{equation}
S'_0(r) = \frac{1}{\ell} \left[\frac{I_1(r/\ell)}{I_2(R_0/\ell)} - \frac{2\ell}{r}\frac{I_2(r/\ell)}{I_2(R_0/\ell)}\right].
\end{equation}

In Fourier space, the boundary conditions \cref{eq bc-theta,eq bc-S} read
\begin{subequations} \label{eq bc-Fourier}
\begin{align}
&\delta\tilde{\theta}_k(R_0) = -\frac{ik}{R_0}\delta\tilde{R}_k, \label{eq bc-Fourier-theta}\\
&\delta\tilde{S}_k(R_0) = -S'_0(R_0)\delta\tilde{R}_k, \label{eq bc-Fourier-S}
\end{align}
\end{subequations}
where $\delta\tilde{R}_k$ are the angular Fourier components of the radius perturbation $\delta R(\phi)$. Applying these boundary conditions to the solutions \cref{eq perturbed-S-ansatz,eq perturbed-theta-ansatz}, I obtain the final solutions
\begin{subequations} \label{eq nematic-perturbations}
\begin{multline}
\delta\tilde{S}_k(r) = \left\{ \left[\frac{1+k}{R_0} - \frac{1}{2\ell}\frac{I_1(R_0/\ell)}{I_2(R_0/\ell)}\right]\frac{I_{k+2}(r/\ell)}{I_{k+2}(R_0/\ell)} \right.\\
\left.+ \left[\frac{1-k}{R_0} - \frac{1}{2\ell} \frac{I_1(R_0/\ell)}{I_2(R_0/\ell)}\right] \frac{I_{k-2}(r/\ell)}{I_{k-2}(R_0/\ell)} \right\} \delta\tilde{R}_k,
\end{multline}
\begin{multline}
\delta\tilde{\theta}_k(r) = \left\{ \left[\frac{1-k}{R_0} - \frac{1}{2\ell} \frac{I_1(R_0/\ell)}{I_2(R_0/\ell)}\right] \frac{I_{k-2}(r/\ell)}{I_{k-2}(R_0/\ell)} \right.\\
\left.- \left[\frac{1+k}{R_0} - \frac{1}{2\ell}\frac{I_1(R_0/\ell)}{I_2(R_0/\ell)}\right]\frac{I_{k+2}(r/\ell)}{I_{k+2}(R_0/\ell)} \right\} \frac{I_2(R_0/\ell)}{I_2(r/\ell)} \frac{i}{2}\delta\tilde{R}_k.
\end{multline}
\end{subequations}

\subsection{Forces and flows} \label{perturbed-flow}

The perturbations of the nematic order obtained above induce flows that further affect droplet shape. To obtain these flows, we first compute the pressure perturbations. The perturbations of the active pressure source in \cref{eq pressure} read as
\begin{multline}
\delta s(r,\phi) = - \zeta \cos(2\theta_{\text{a}}) \left\{ \left[\partial_r^2 +\frac{3}{r} \partial_r -\frac{1}{r^2}\partial_\phi^2\right] \delta S(r,\phi) \right.\\
\left.+ \frac{4}{r}\left\{ S_0(r)\left[\frac{1}{r} +\partial_r\right] +S'_0(r)\right\} \partial_\phi \delta\theta(r,\phi)\right\} \\
- 2\zeta \sin(2\theta_{\text{a}}) \left\{ \left[ \frac{1}{r}\partial_r + \frac{1}{r^2}\right] \partial_\phi \delta S(r,\phi) \right.\\
- \left\{ S_0(r) \partial_r^2 + \left[\frac{3}{r} S_0(r) + 2 S'_0(r)\right] \partial_r \right.\\
\left. \left. + \frac{3}{r} S'_0(r) + S''_0(r) - \frac{1}{r^2} S_0(r) \partial_\phi^2 \right\} \delta\theta(r,\phi) \right\}.
\end{multline}
Its angular Fourier components are given by
\begin{multline} \label{eq pressure-source-perturbations}
\delta\tilde s_k(r) = -\zeta \cos(2\theta_{\text{a}}) \left\{\left[ \frac{\dd^2}{\dd r^2} + \frac{3}{r}\frac{\dd}{\dd r} + \frac{k^2}{r^2}\right] \delta\tilde S_k(r) \right.\\
\left. + \frac{4}{r}\left\{ S_0(r) \left[ \frac{1}{r} + \frac{\dd}{\dd r}\right] + S'_0(r) \right\} i k\, \delta\tilde \theta_k(r)\right\} \\
- 2\zeta \sin(2\theta_{\text{a}}) \left\{ \left[\frac{1}{r}\frac{\dd}{\dd r} + \frac{1}{r^2} \right] ik\, \delta\tilde S_k(r) \right.\\
- \left\{S_0(r) \frac{\dd^2}{\dd r^2} + \left[\frac{3}{r}S_0(r) + 2 S'_0(r) \right] \frac{\dd}{\dd r} \right.\\
\left. \left. + \frac{3}{r} S'_0(r) + S''_0(r) + \frac{k^2}{r^2} S_0(r) \right\} \delta\tilde\theta_k(r)\right\}.
\end{multline}
In Fourier space, the Poisson equation \cref{eq pressure} can be recast as
\begin{equation} \label{eq Poisson-perturbations}
\left[ \frac{\dd^2}{\dd r^2} + \frac{1}{r}\frac{\dd}{\dd r} - \frac{k^2}{r^2}\right] \delta\tilde P_k(r) = \delta\tilde s_k (r).
\end{equation}
Then, introducing the nematic order perturbations \cref{eq nematic-perturbations} into the pressure source perturbations \cref{eq pressure-source-perturbations}, I solve \cref{eq Poisson-perturbations} to obtain the angular Fourier components of the pressure perturbations. For $k>0$, they are given by
\begin{multline} \label{eq pressure-perturbations-general}
\delta\tilde P_k(r) = A_k r^k + \frac{B_k}{r^k} \\
-\zeta \frac{\delta\tilde R_k}{R_0} c_k(R_0/\ell,\theta_{\text{a}}) \left[ \frac{1}{k!}\left(\frac{r}{2\ell}\right)^k - I_k(r/\ell)\right],
\end{multline}
where $A_k$ and $B_k$ are integration constants, and
\begin{multline}
c_k = \frac{(2\ell/R_0)^{k+1} (k+2)!}{_0F_1(k+3; R_0^2/(4\ell^2))} \\
\times e^{-2i\theta_{\text{a}}} \left\{ \left[ 1 + e^{4i\theta_{\text{a}}} + 4k(k+1) \frac{\ell^2}{R_0^2} \right.\right.\\
\left. - 4k\left(1 + 2 (k^2-1)\frac{\ell^2}{R_0^2}\right)\frac{\ell}{R_0} \frac{I_{k-1}(R_0/\ell)}{I_{k-2}(R_0/\ell)}\right] \frac{I_1(R_0/\ell)}{I_2(R_0/\ell)}\\
+ \left[ k-1 -(k+1) e^{4i\theta_{\text{a}}} + 4k(k^2-1) \frac{\ell^2}{R_0^2} \right.\\
\left.\left.- 4k(k-1) \left(1 + 2 (k^2-1)\frac{\ell^2}{R_0^2}\right)\frac{\ell}{R_0} \frac{I_{k-1}(R_0/\ell)}{I_{k-2}(R_0/\ell)}\right] \frac{2\ell}{R_0} \right\}
\end{multline}
is a numerical factor. Here, $_0F_1$ is a generalized hypergeometric function. For mode $k=0$, the solution for the pressure perturbation is different:
\begin{equation}
\delta\tilde P_0(r) = A_0 + B_0 \ln r - \zeta \frac{\delta\tilde R_k}{R_0} c_0(R_0/\ell,\theta_{\text{a}}) \left[I_0(r/\ell)-1\right].
\end{equation}

The integration constants $A_k$ and $B_k$ in \cref{eq pressure-perturbations-general} are determined by the Young-Laplace boundary condition \cref{eq Young-Laplace}. To first order in perturbations, it implies
\begin{equation} \label{eq stress-boundary}
\sigma_{rr}(R) = -\frac{\gamma}{R_0} \left\{ 1 - \left[ 1+ \frac{\dd^2}{\dd \phi^2}\right] \frac{\delta R}{R_0}\right\},
\end{equation}
where we have used that $\sigma_{r\phi}^0 = 0$. Then, taking into account that $\sigma_{rr}(R) \approx \sigma_{rr}^0 (R) + \delta\sigma_{rr}(R) \approx \sigma^0_{rr} (R_0) + \left.\dd\sigma_{rr}^0/\dd r\right|_{r=R_0}\delta R(\phi) + \delta\sigma_{rr}(R_0)$, as well as that $\sigma^0_{rr} \approx -P_0 - \zeta \cos(2\theta_{\text{a}}) S_0$ and $\delta\sigma_{rr} = -\delta P - \zeta \cos(2\theta_{\text{a}}) \delta S - 2\zeta S_0 \sin(2\theta_{\text{a}}) \delta\theta$, \cref{eq stress-boundary} translates into a boundary condition for the pressure perturbations
\begin{multline}
- \left[P'_0(R_0) + \zeta \cos(2\theta_{\text{a}}) S'_0(R_0)\right] \delta R(\phi) - \delta P(R_0,\phi)\\
 - \zeta \cos(2\theta_{\text{a}}) \delta S(R_0,\phi) - 2\zeta \sin(2\theta_{\text{a}}) S_0(R_0) \delta\theta(R_0,\phi) \\
 \approx \frac{\gamma}{R_0^2} \left[1 + \frac{\dd^2}{\dd \phi^2}\right] \delta R.
\end{multline}
In Fourier space, this condition reads as
\begin{multline} \label{eq Young-Laplace-Fourier}
\delta\tilde P_k(R_0) = \frac{\gamma}{R_0^2} (k^2-1) \delta\tilde R_k \\
- \left[ P'_0(R_0) + \zeta \cos(2\theta_{\text{a}}) S'_0(R_0)\right] \delta\tilde R_k\\
 - \zeta \cos(2\theta_{\text{a}}) \delta\tilde S_k(R_0) - 2\zeta \sin(2\theta_{\text{a}}) \delta\tilde\theta_k(R_0).
\end{multline}
For all modes, this boundary condition determines the integration constant $A_k$, whereas $B_k$ must vanish to avoid the pressure field to diverge at $r\rightarrow 0$. Introducing the values of these integration constants yields the final solutions for the pressure perturbation modes $\delta\tilde P_k(r)$.

Next, we can obtain the flow perturbations by means of the force balance \cref{eq force-balance}. For the radial velocity perturbations, it implies
\begin{equation}
\xi \delta v_r = - \partial_r \delta P + \delta f_r^{\text{a}}.
\end{equation}
Therefore, the angular Fourier components of the radial velocity perturbations are given by
\begin{equation}
\delta\tilde v_{r,k}(r) = \frac{1}{\xi}\left[ -\frac{\dd \delta\tilde P_k(r)}{\dd r} + \delta\tilde f_{r,k}^{\text{a}}(r)\right],
\end{equation}
where
\begin{multline}
\delta\tilde f_{r,k}^{\text{a}}(r) \\
= -\zeta \cos(2\theta_{\text{a}}) \left\{ \left[ \frac{\dd}{\dd r} + \frac{2}{r}\right] \delta\tilde S_k(r) + \frac{2ik}{r}S_0(r) \delta\tilde\theta_k(r)\right\}\\
-\zeta \sin(2\theta_{\text{a}}) \left\{ \frac{ik}{r}\delta\tilde S_k(r) \right.\\
\left. - 2 \left[ S_0(r) \frac{\dd}{\dd r} + S'_0(r) + \frac{2}{r}S_0(r)\right] \delta\tilde\theta_k(r) \right\}.
\end{multline}

\subsection{Interface dynamics} \label{interface-dynamics}

The dynamics of the droplet interface is given by the free-boundary kinematic condition in \cref{eq kinematic-condition}. In Fourier space, it reads as
\begin{equation}
\frac{\dd \delta\tilde{R}_k}{\dd t} \approx \delta\tilde{v}_{r,k}(R_0).
\end{equation}
Thus, the linear growth rate $\omega_k$ of the radius perturbations, defined by $\dd \delta\tilde{R}_k/\dd t = \omega_k \delta\tilde{R}_k$ is given by
\begin{equation}
\omega_k = \frac{\delta\tilde{v}_{r,k}(R_0)}{\delta\tilde{R}_k}.
\end{equation}

Introducing all previous results, I obtain the final result:
\begin{equation}
\omega_k = \frac{e^{-2i \theta_{\text{a}}}}{2\ell^2 \xi} \left[ \zeta a_k(R_0/\ell, \theta_{\text{a}}) - \frac{\gamma}{R_0} b_k(R_0/\ell, \theta_{\text{a}})\right],
\end{equation}
where the factors $a_k$ and $b_k$ are given by
\begin{widetext}
\begin{multline}
a_k(R_0/\ell, \theta_{\text{a}}) = \frac{ I_{k-1}(R_0/\ell) }{ I_2(R_0/\ell) I_{k-2}(R_0/\ell) I_{k+2}(R_0/\ell) } \left[ I_1(R_0/\ell) + 2(k-1) \frac{\ell}{R_0} I_2(R_0/\ell) \right] \\
\times \left[ 4k\frac{\ell}{R_0} \left[ 1 + 2 (k^2-1) \frac{\ell^2}{R_0^2} \right] I_{k+1}(R_0/\ell) + I_{k+2}(R_0/\ell) \right] \\
+ \frac{4 \ell^4}{R_0^4} \frac{1}{ _0F_1(k+3; R_0^2/(4\ell^2)) } \left\{ 2 k \left(3k-2 - k e^{4i\theta_{\text{a}}} \right)  \,  _0F_1(k+1; R_0^2/(4\ell^2)) - 4(k-1) \, _0F_1(k; R_0^2/(4\ell^2)) \phantom{\frac{ _0F_1(2; R_0^2/(4\ell^2)) }{ _0F_1(3; R_0^2/(4\ell^2)) }} \right.\\
 -8 k (k+1) \frac{\ell^2}{R_0^2} \left(3k-2 - k e^{4i\theta_{\text{a}}} \right) \left[ \, _0F_1(k; R_0^2/(4\ell^2)) - \, _0F_1(k+1; R_0^2/(4\ell^2)) \right] \\
\left. + \frac{ _0F_1(2; R_0^2/(4\ell^2)) }{ _0F_1(3; R_0^2/(4\ell^2)) } \left[ \frac{1}{2} (e^{4i\theta_{\text{a}}} - 1) \left[ \frac{R_0^2}{\ell^2} + 4k (k+1) \right] \, _0F_1(k+2; R_0^2/(4\ell^2)) - 2 (e^{4i\theta_{\text{a}}} + 1) \, _0F_1(k; R_0^2/(4\ell^2)) \right] \right\},
\end{multline}
\begin{multline}
b_k(R_0/\ell, \theta_{\text{a}}) = \frac{8 \ell^4}{R_0^4} \frac{k (k^2-1) e^{2i\theta_{\text{a}}} }{ _0F_1(k+3; R_0^2/(4\ell^2)) } \left\{ \, _0F_1(k+1; R_0^2/(4\ell^2)) \phantom{\frac{\ell^2}{R_0^2}} \right.\\
\left. - 4 \frac{\ell^2}{R_0^2} (k+1) \left[ \, _0F_1(k; R_0^2/(4\ell^2)) - \, _0F_1(k+1; R_0^2/(4\ell^2)) \right] \right\}.
\end{multline}
\end{widetext}


\end{document}